\documentclass[twocolumn,prl,a4paper,superscriptaddress,showpacs,preprintnumbers,tightenlines]{revtex4}
\usepackage{graphicx} 
\usepackage{dcolumn}  
\usepackage{epsfig}
\usepackage{amssymb}

\graphicspath{{figs/}}

\begin{document}

\title{ \quad\\[0.5cm] \boldmath
  Evidence of $B^0 \to \rho^0 \pi^0$
}

\affiliation{Budker Institute of Nuclear Physics, Novosibirsk}
\affiliation{Chiba University, Chiba}
\affiliation{Chonnam National University, Kwangju}
\affiliation{University of Cincinnati, Cincinnati, Ohio 45221}
\affiliation{Gyeongsang National University, Chinju}
\affiliation{University of Hawaii, Honolulu, Hawaii 96822}
\affiliation{High Energy Accelerator Research Organization (KEK), Tsukuba}
\affiliation{Hiroshima Institute of Technology, Hiroshima}
\affiliation{Institute of High Energy Physics, Chinese Academy of Sciences, Beijing}
\affiliation{Institute of High Energy Physics, Vienna}
\affiliation{Institute for Theoretical and Experimental Physics, Moscow}
\affiliation{J. Stefan Institute, Ljubljana}
\affiliation{Kanagawa University, Yokohama}
\affiliation{Korea University, Seoul}
\affiliation{Kyungpook National University, Taegu}
\affiliation{Swiss Federal Institute of Technology of Lausanne, EPFL, Lausanne}
\affiliation{University of Ljubljana, Ljubljana}
\affiliation{University of Maribor, Maribor}
\affiliation{University of Melbourne, Victoria}
\affiliation{Nagoya University, Nagoya}
\affiliation{Nara Women's University, Nara}
\affiliation{National United University, Miao Li}
\affiliation{Department of Physics, National Taiwan University, Taipei}
\affiliation{H. Niewodniczanski Institute of Nuclear Physics, Krakow}
\affiliation{Nihon Dental College, Niigata}
\affiliation{Niigata University, Niigata}
\affiliation{Osaka City University, Osaka}
\affiliation{Osaka University, Osaka}
\affiliation{Panjab University, Chandigarh}
\affiliation{Peking University, Beijing}
\affiliation{Princeton University, Princeton, New Jersey 08545}
\affiliation{University of Science and Technology of China, Hefei}
\affiliation{Seoul National University, Seoul}
\affiliation{Sungkyunkwan University, Suwon}
\affiliation{University of Sydney, Sydney NSW}
\affiliation{Tata Institute of Fundamental Research, Bombay}
\affiliation{Toho University, Funabashi}
\affiliation{Tohoku Gakuin University, Tagajo}
\affiliation{Tohoku University, Sendai}
\affiliation{Department of Physics, University of Tokyo, Tokyo}
\affiliation{Tokyo Institute of Technology, Tokyo}
\affiliation{Tokyo Metropolitan University, Tokyo}
\affiliation{Tokyo University of Agriculture and Technology, Tokyo}
\affiliation{University of Tsukuba, Tsukuba}
\affiliation{Virginia Polytechnic Institute and State University, Blacksburg, Virginia 24061}
\affiliation{Yonsei University, Seoul}
  \author{J.~Dragic}\affiliation{University of Melbourne, Victoria} % Melbourne
  \author{T.~Gershon}\affiliation{High Energy Accelerator Research Organization (KEK), Tsukuba} % KEK
  \author{K.~Abe}\affiliation{High Energy Accelerator Research Organization (KEK), Tsukuba} % KEK
  \author{K.~Abe}\affiliation{Tohoku Gakuin University, Tagajo} % TohokuGakuin
  \author{T.~Abe}\affiliation{High Energy Accelerator Research Organization (KEK), Tsukuba} % KEK
  \author{H.~Aihara}\affiliation{Department of Physics, University of Tokyo, Tokyo} % Tokyo
  \author{Y.~Asano}\affiliation{University of Tsukuba, Tsukuba} % Tsukuba
  \author{V.~Aulchenko}\affiliation{Budker Institute of Nuclear Physics, Novosibirsk} % BINP
  \author{T.~Aushev}\affiliation{Institute for Theoretical and Experimental Physics, Moscow} % ITEP
  \author{T.~Aziz}\affiliation{Tata Institute of Fundamental Research, Bombay} % Tata
% ...
% Ljubljana
\collaboration{The Belle Collaboration}

\begin{abstract}
  \noindent
  We present the first evidence of the decay $B^0 \to \rho^0 \pi^0$, using
  $140\ {\rm fb}^{-1}$ of data collected at the $\Upsilon(4S)$
  resonance with the Belle detector at the KEKB asymmetric $e^+e^-$
  collider. We detect a signal 
  with a significance of 3.5 standard deviations, and measure the branching
  fraction to be ${\mathcal B}\left( B^0 \to \rho^0\pi^0 \right) =
  \left( 5.1 \pm 1.6 ({\rm stat}) \pm 0.9 ({\rm syst}) \right)
  \times 10^{-6}$.
\end{abstract}


\maketitle

\tighten

{\renewcommand{\thefootnote}{\fnsymbol{footnote}}}
\setcounter{footnote}{0}
Recent measurements of the $CP$ violating parameter
$\sin 2\phi_1$~\cite{ref:belle_sin2phi1,ref:babar_sin2phi1}
have confirmed the Kobayashi-Maskawa mechanism~\cite{ref:km} as the origin
of $CP$ violation within the Standard Model (SM).
It is now essential to test the SM via measurements of
other $CP$ violating parameters.
Of particular importance are the other two angles
of the Unitarity Triangle, $\phi_2$ and $\phi_3$.
Measurements of $\phi_2$ typically rely on time-dependent studies of
decays of $B$ mesons to light mesons,
such as $B^0 \to \pi^+\pi^-$ and $\rho^{\pm}\pi^{\mp}$.
Although these analyses are complicated by the presence of penguin amplitudes,
isospin analyses can be used to extract $\phi_2$~\cite{ref:isospin}.
Recent evidence for direct $CP$ violation
in $B^0 \to \pi^+\pi^-$~\cite{ref:belle_pipi}
indicates a sizeable penguin contribution;
furthermore measurements of the $B^0 \to \pi^0\pi^0$ branching fraction
at a level higher than most theoretical expectations~\cite{ref:pi0pi0}
suggest that much larger data samples will be needed for a
model-independent
extraction of $\phi_2$ from the $\pi\pi$ system using an isospin analysis.

Measurements of $\phi_2$ from the $\rho\pi$ system
rely on knowledge of the branching fraction of $B^0 \to \rho^0\pi^0$.
The isospin analysis depends on this information,
along with the $CP$ asymmetry,
since all the other $\rho\pi$ final states have been
observed~\cite{ref:rhopi_bfs,ref:rho0pi0_ul}.
An alternative technique to extract $\phi_2$ uses an amplitude analysis
of $B^0 \to \pi^+\pi^-\pi^0$~\cite{ref:rhopi_dalitz}.
Since $B^0 \to \rho^0\pi^0$ results in this final state, it is essential
to understand its contribution, as well as possible effects
from scalar resonances, {\it e.g.} $\sigma\pi^0$,
and nonresonant sources~\cite{ref:sigmapi0}.

Recent theoretical predictions for the branching fraction of
$B^0 \to \rho^0 \pi^0$ are typically around or below
$10^{-6}$~\cite{ref:rho0pi0_predictions},
while the most restrictive experimental upper limit,
recently set by the BaBar Collaboration, is
${\cal B}\left( B^0 \to \rho^0 \pi^0 \right) < 2.9 \times 10^{-6}$~\cite{ref:rho0pi0_ul}
at the $90\%$ confidence level.
In this letter, we present the first evidence for $B^0 \to \rho^0 \pi^0$.

The analysis is based on a $140\ {\rm fb}^{-1}$ data sample 
containing $152 \times 10^6$ $B$ meson pairs collected
with the Belle detector at
the KEKB asymmetric-energy $e^+e^-$ collider~\cite{ref:KEKB}. 
KEKB operates at the $\Upsilon(4S)$ resonance 
($\sqrt{s}=10.58~{\rm GeV}$) with
a peak luminosity that exceeds
$1.2 \times 10^{34}~{\rm cm}^{-2}{\rm s}^{-1}$.
The production rates of $B^+B^-$ and $B^0\bar{B}^0$ pairs are assumed
to be equal.

The Belle detector is a large-solid-angle magnetic spectrometer that
consists of a three-layer silicon vertex detector (SVD),
a 50-layer central drift chamber (CDC), 
an array of aerogel threshold \v{C}erenkov counters (ACC), 
a barrel-like arrangement of time-of-flight scintillation counters (TOF), 
and an electromagnetic calorimeter comprised of CsI(Tl) crystals (ECL) 
located inside a superconducting solenoid coil 
that provides a $1.5\ {\rm T}$ magnetic field.  
An iron flux-return located outside of the coil is instrumented 
to detect $K_L$ mesons and to identify muons (KLM).  
The detector is described in detail elsewhere~\cite{ref:belle}.

Charged tracks are required to originate from the interaction point 
and have transverse momenta greater than $100~{\rm MeV}/c$. 
To identify tracks as charged pions,
we combine specific ionisation measurements from the CDC,
pulse height information from the ACC and 
timing information from the TOF into
pion/kaon likelihood variables ${\mathcal L}_{\pi/K}$.
We then require
${\mathcal L}_{\pi}/\left( {\mathcal L}_{\pi} + {\mathcal L}_K \right) > 0.6$,
which provides a pion selection efficiency of 93\% while keeping the
kaon misidentification probability below 10\%.
Additionally, we reject tracks that are consistent with an electron hypothesis.

Neutral pion candidates are reconstructed from photon pairs
with invariant masses in the range 
$0.115\ {\rm GeV}/c^2 < m_{\gamma\gamma} < 0.154\ {\rm GeV}/c^2$,
corresponding to a window of $\pm3\sigma$ about the nominal $\pi^0$ mass,
where $\sigma$ is the experimental resolution 
for the most energetic $\pi^0$ candidates.
Photon candidates are selected with a minimum energy requirement 
of $50~{\rm MeV}$ in the barrel region of the ECL,
defined as $32^\circ < \theta_\gamma < 129^\circ$
and $100~{\rm MeV}$ in the endcap regions,
defined as $17^\circ < \theta_\gamma < 32^\circ$ and
$129^\circ < \theta_\gamma < 150^\circ$, 
where $\theta_\gamma$ denotes the polar angle of the photon with 
respect to the beam line.
The $\pi^0$ candidates are required to have transverse momenta 
greater than $100~{\rm MeV}/c$ in the laboratory frame.
In addition, we make a loose requirement on 
the goodness of fit of a $\pi^0$ mass-constrained fit of
$\gamma\gamma$ ($\chi^2_{\pi^0}$).

Possible contributions to the $\pi^+\pi^-\pi^0$ final state
from charmed ($b \to c$) backgrounds are  explicitly vetoed
for the decays $B^0 \to D^- \pi^+$, $\bar{D}^0 \pi^0$ and $J/\psi \pi^0$,
based on the two-particle invariant masses. 
From Monte Carlo (MC) simulation, we find a small
combinatorial background from $b \to c$ remains.

$B$ candidates are selected using two kinematic variables:
the beam-constrained mass
$M_{\rm bc}\equiv \sqrt{E^2_{\rm beam}-p^2_B}$ 
and the energy difference $\Delta E \equiv E_B - E_{\rm beam}$.
Here, $E_B$ and $p_B$ are the reconstructed energy and momentum 
of the $B$ candidate in the centre of mass (CM) frame, 
and $E_{\rm beam}$ is the beam energy in the CM frame.
We consider candidate events in the region 
$-0.20\ {\rm GeV} < \Delta E < 0.40\ {\rm GeV}$ and
$ 5.23\ {\rm GeV}/c^2 < M_{\rm bc} < 5.30\ {\rm GeV}/c^2$.
With these boundaries 30\% of events have more than one candidate,
and that with the smallest $\chi^2_{\rm vtx} + \chi^2_{\pi^0}$ is selected,
where $\chi^2_{\rm vtx}$ is 
the goodness of fit of a vertex-constrained fit of $\pi^+\pi^-$.
We define signal regions in $\Delta E$ and $M_{\rm bc}$ as
$-0.135\ {\rm GeV} < \Delta E < 0.082\ {\rm GeV}$ and
$5.269\ {\rm GeV}/c^2 < M_{\rm bc} < 5.290\ {\rm GeV}/c^2$ 
respectively.

To select $\rho^0\pi^0$ from the three-body 
$\pi^+\pi^-\pi^0$ candidates, we require
the $\pi^+\pi^-$ invariant mass to be in the range
$0.50\ {\rm GeV}/c^2 < m_{\pi^+\pi^-} < 1.10\ {\rm GeV}/c^2$ and 
the $\rho^0$ helicity angle to satisfy
$\left| \cos \theta^{\rho}_{\rm hel} \right| > 0.5$,
where $\theta^{\rho}_{\rm hel}$ is 
defined as the angle between the negative pion direction and the
opposite of the $B$ direction in the $\rho$ rest frame~\cite{helicity}.
Contributions from $B^0 \to \rho^{\pm}\pi^{\mp}$ 
are explicitly vetoed by rejecting candidates 
with $\pi^\pm\pi^0$ invariant masses 
that fall into the window
$0.50\ {\rm GeV}/c^2 < m_{\pi^{\pm}\pi^0} < 1.10\ {\rm GeV}/c^2$.
This requirement also vetoes the region of the Dalitz plot
where the interference between $\rho$ resonances is strongest.

The dominant background to $B^0 \to \pi^+\pi^-\pi^0$
comes from continuum events, $e^+e^-\to q\bar{q}$ ($q = u, d, s, c$).
Since these tend to be jet-like, 
whilst $B\bar{B}$ events tend to be spherical,
we use event shape variables to discriminate between the two.
We combine five modified Fox-Wolfram moments~\cite{fox-wolfram}
into a Fisher discriminant and tune the 
coefficients to maximise the separation between
signal and continuum events.
We define $\theta_B$ as the angle of the reconstructed $B$
candidate with respect to the beam direction in the CM frame.
Signal events have a distribution proportional to $\sin^2\theta_B$,
whilst continuum events are flatly distributed in $\cos\theta_B$.
We combine the output of the Fisher discriminant 
with $\cos\theta_B$ into signal/background likelihood variables, 
${\mathcal L}_{s/b}$, and define the likelihood ratio
${\mathcal R}={\mathcal L}_s/\left({\mathcal L}_s + {\mathcal L}_b \right)$.
In order to maximise the separation between signal and background,
we make use of the additional discriminatory power provided by
the flavour tagging algorithm developed for time-dependent analyses
at Belle~\cite{Kakuno:2004cf}.
We utilise the parameter $r$, which takes values between 0 and 1 and can
be used as a measure of the confidence that the remaining particles in the
event (other than $\pi^+\pi^-\pi^0$) originate from a flavour specific $B$
meson decay.
Events with a high value of $r$ are considered well-tagged 
and hence are unlikely to have originated from continuum processes.
Moreover, we find that there is no strong correlation with 
any of the topological variables used above to separate signal from continuum.

\begin{figure}[h!]
  \hbox to \hsize{
    \hss
    \includegraphics[width=0.95\hsize]{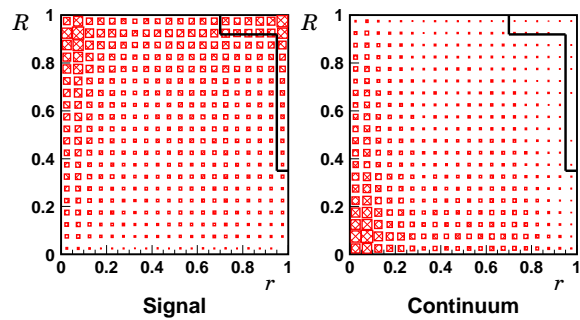}
    \hss
  }
  \caption{Distributions of signal (MC) and continuum (sideband data)
    events in $r$-${\cal R}$
    space. The marked region indicates the selection requirement
    obtained from the optimisation procedure described in the text.
  }
  \label{fig:r-R_space}
\end{figure}

We use a continuum suppression requirement on $r$ and ${\cal R}$ that 
maximises the value of $N_s/\sqrt{N_s+N_b}$, 
where $N_s$ and $N_b$ are the numbers of signal and background events 
contained in the intersection of the $\Delta E$ and $M_{\rm bc}$ signal areas.
To obtain $N_s$ we use a large statistics sample of $\rho^0\pi^0$ 
MC, and assume a branching fraction for
$B^0\to \rho^0\pi^0$ of $1 \times 10^{-6}$.
We estimate $N_b$ from a continuum dominated sideband in data,
defined as the union of the two regions
$-0.20\ {\rm GeV} < \Delta E <0.40\ {\rm GeV}$ and 
$5.23\ {\rm GeV}/c^2 < M_{\rm bc} < 5.26\ {\rm GeV}/c^2$, and
$0.20\ {\rm GeV} < \Delta E < 0.40\ {\rm GeV}$ and 
$5.26\ {\rm GeV}/c^2 < M_{\rm bc} < 5.30\ {\rm GeV}/c^2$.
We use an iterative procedure to find the optimal contiguous area in
$r$-${\cal R}$ space. This method is found to be robust against statistical
fluctuations in the samples used to obtain $N_s$ and $N_b$.
The result of the procedure is that we select events that
satisfy either  ${\cal R} > 0.92$ and $r > 0.70$ or
${\cal R} > 0.35$ and $r > 0.95$, as
shown in Fig.~\ref{fig:r-R_space}. In addition to optimising
$N_s/\sqrt{N_s+N_b}$, this requirement is found to improve
$N_s/N_b$ by a factor of 76.

\begin{figure*}
  \hbox to \hsize{
    \hss
    \includegraphics[width=0.15\vsize]{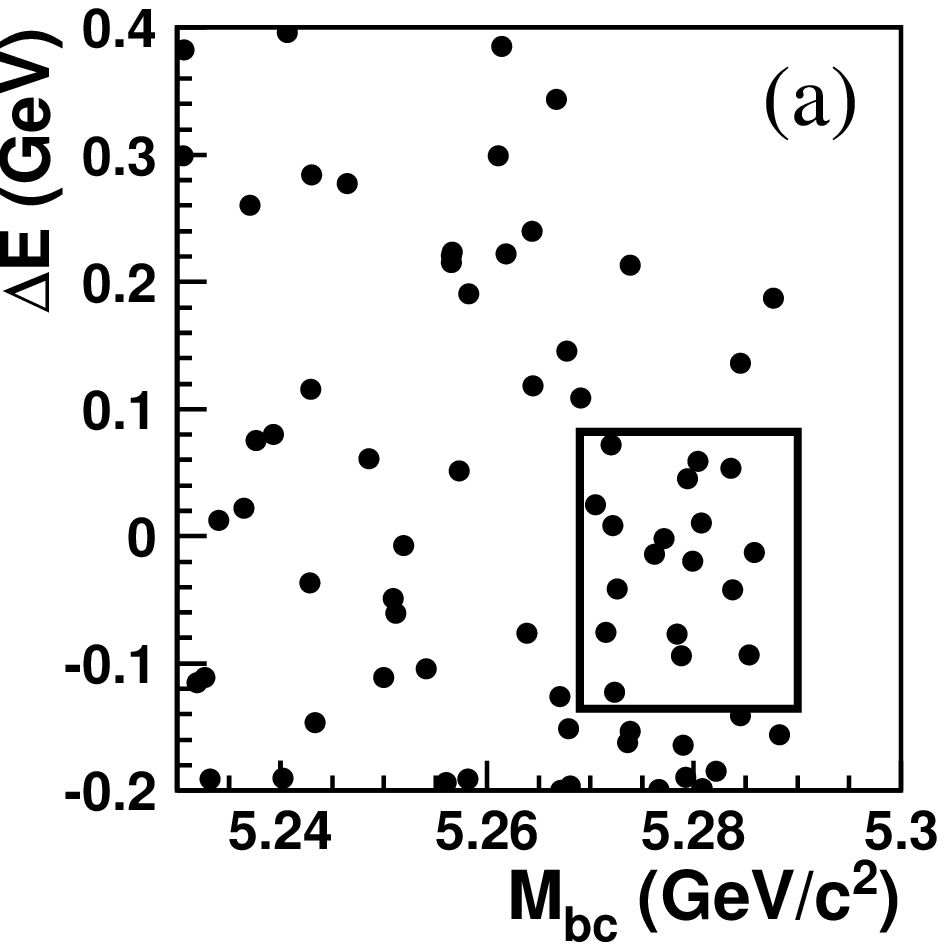}\hspace{-1.5ex}
    \includegraphics[width=0.4\hsize]{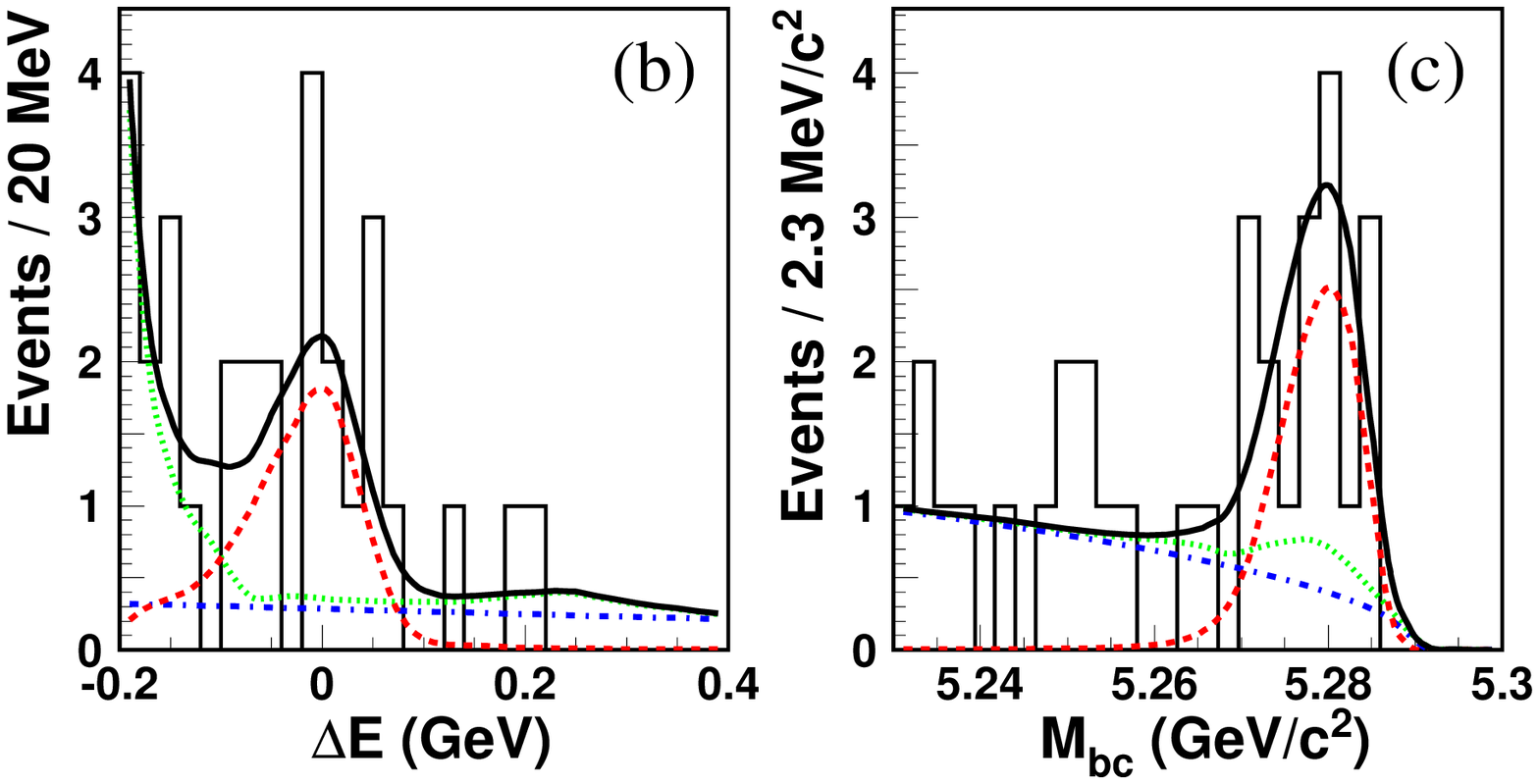}\hspace{-1.5ex}
    \includegraphics[width=0.4\hsize]{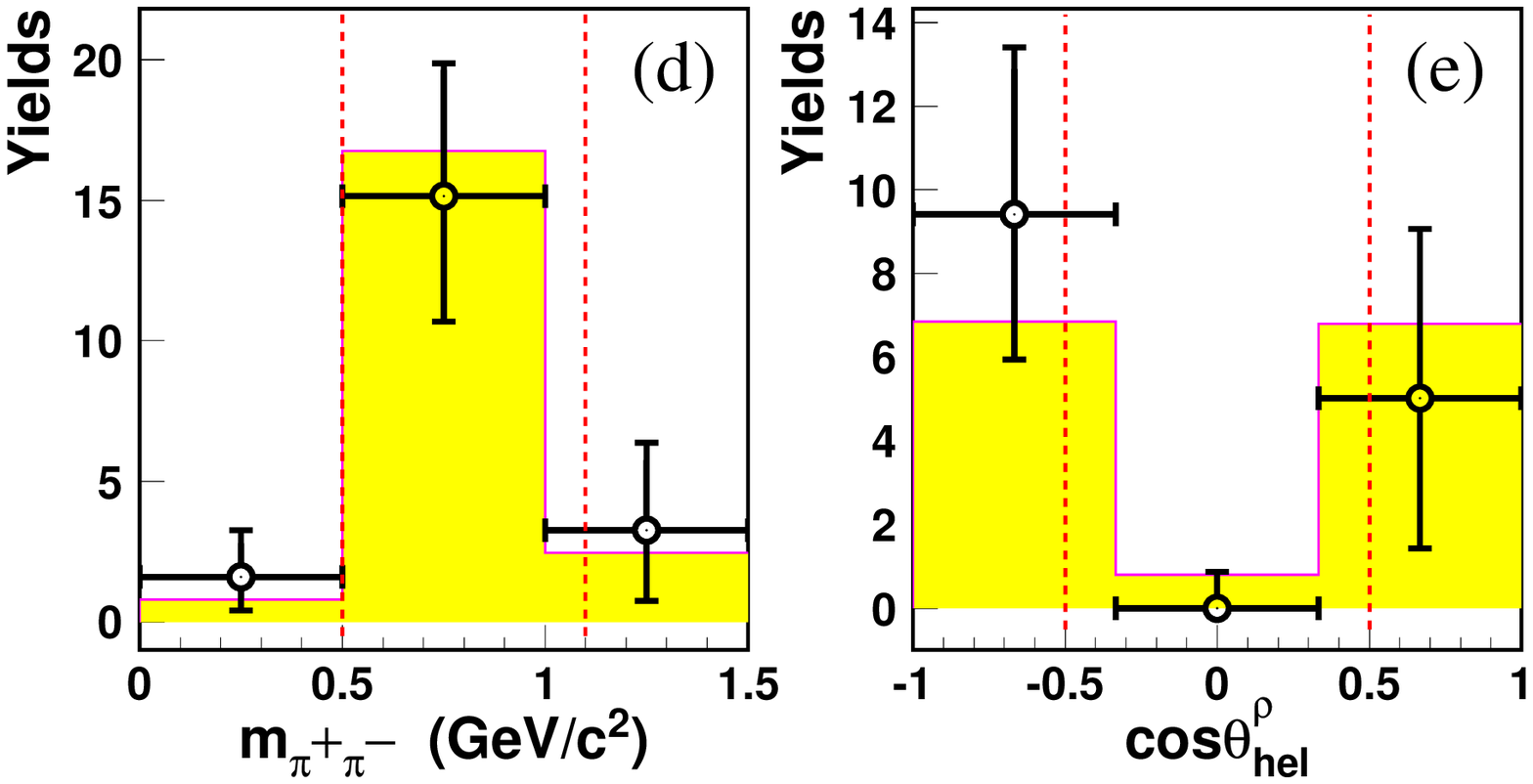}
    \hss
  }
  \caption{ (a) Scatter plot of $\Delta E$ versus $M_{\rm bc}$
    for the selected candidate events; the box indicates the intersection of
    $\Delta E$ and $M_{\rm bc}$ signal regions.
    (b), (c) Distribution of $\Delta E$($M_{\rm bc}$) in the signal
    region of $M_{\rm bc}$($\Delta E$).
    Projection of the fit result is shown as the solid curve;
    the dashed line represents the signal component;
    the dot-dashed curve represents the contribution from continuum events,
    and the dotted curve represents the composite of continuum
    and $B$-related backgrounds. (d), (e)  Distributions of fit yields
    in $m_{\pi^+\pi^-}$ and $\cos \theta^{\rho}_{\rm hel}$ variables
    for $\rho^0\pi^0$ candidate events. 
    Points with error bars represent data fit results, and the histograms
    show signal MC expectation; the selection requirements described
    in the text are shown as dashed lines.  
  }
  \label{fig:rho0pi0_fit}
\end{figure*}

Following all the selection criteria described above, the signal
efficiency measured in MC is found to be $(1.91 \pm 0.01)\%$, and we
find 73 candidates remain in the data,
as shown in Fig.~\ref{fig:rho0pi0_fit}(a).
We obtain the signal yield using an unbinned maximum-likelihood fit
to the $\Delta E$-$M_{\rm bc}$ distribution
of the selected candidate events.  
The fitting function contains components for the signal, 
continuum background, $b \to c$ background 
and the charmless $B$ decays $B^+ \to \rho^+ \rho^0$,
$B^+ \to \rho^+ \pi^0$ and $B^+ \to \pi^+ \pi^0$. 
The possible contribution from other charmless $B$ decays 
is found to be small (0.7 events) using a large MC sample~\cite{rareb},
and is taken into account in the systematic error.
The probability density functions (PDFs) for the 
signal and charmless $B$ backgrounds are taken from 
smoothed two dimensional histograms obtained from large MC samples.
For $B^+ \to \rho^+ \rho^0$ our MC assumes $100\%$
longitudinal polarisation~\cite{ref:rhoprho0_bf}.
For the signal PDF, small corrections to MC peak positions
($ < 0.5$ MeV) and widths ($< 16\%$) are applied. These factors 
are derived from control samples
($B^0 \to D^{*-}\rho^+$ with
$D^{*-} \to \bar{D}^0\pi^-$, $\bar{D}^0 \to K^+\pi^-$, $\rho^+ \to \pi^+\pi^0$ 
and
$B^+ \to \bar{D}^0\rho^+$ with 
$\bar{D}^0 \to K^+\pi^-$, $\rho^+ \to \pi^+\pi^0$),
in which we require that the $\pi^0$ momentum be greater than
$1.8\ {\rm GeV}/c$ in order to mimic the high momentum $\pi^0$ in our signal.

The two-dimensional PDF for the continuum background is described as
the product of a first-order polynomial in $\Delta E$  
with an ARGUS function~\cite{argus} in $M_{\rm bc}$.
Contributions from $b \to c$ are also parametrised 
as a product of two one-dimensional PDFs.
Using MC we find the $\Delta E$ distribution of this background
in the fitting region is modeled accurately by an exponential function;
the $M_{\rm bc}$ distribution is modeled by the ARGUS function.
All of the shape parameters describing the continuum and $b \to c$
backgrounds are free parameters in the fit.
The normalisations of $B^+ \to \rho^+\pi^0$ ($2.0 \pm 0.5$ events) and
$B^+ \to \pi^+\pi^0$ ($2.3 \pm 0.5$ events) are fixed in the fit
according to previous measurements~\cite{ref:rho0pi0_ul,rpp_pippiz}, 
while the normalisations of all other components are allowed to float.

The fit result is shown in Fig.~\ref{fig:rho0pi0_fit}(b) and (c).
The signal yield is found to be $15.1 \pm 4.8$ with $3.6 \sigma$ significance.
The significance is defined as 
$\sqrt{-2\ln({\mathcal L}_0/{\mathcal L}_{\rm max})}$, 
where ${\mathcal L}_{\rm max}$ (${\mathcal L}_0$)
denotes the likelihood with the signal yield at its nominal value 
(fixed to zero).
The backgrounds from $b \to c$ and from $B^+ \to \rho^+ \rho^0$
form a peak in the low $\Delta E$ region. The fit results for these
background sources are consistent with the MC expectation, which for
$B^+ \to \rho^+\rho^0$ is based on our branching fraction
measurement~\cite{ref:rhoprho0_bf}.

In order to check that the signal candidates originate from
$B^0 \to \rho^0\pi^0$ decays, 
we change the criteria on $m_{\pi^+\pi^-}$ and $\cos \theta^{\rho}_{\rm hel}$
in turn, and repeat fits to the $\Delta E$-$M_{\rm bc}$ distribution.
The yields obtained in each $m_{\pi^+\pi^-}$ and $\cos \theta^{\rho}_{\rm hel}$
bin are shown in Fig.~\ref{fig:rho0pi0_fit}(d) and (e).

We use the $\cos \theta^{\rho}_{\rm hel}$ distribution to limit contributions
from $\sigma \pi^0$, $f_0(980)\pi^0$ and $\pi^+\pi^-\pi^0$ (nonresonant), 
which are expected to have similar shapes in this variable.
We perform a $\chi^2$ fit including components for 
pseudoscalar $\to$ pseudoscalar vector 
($PV \sim \cos^2 \theta^{\rho}_{\rm hel}$), and
pseudoscalar $\to$ pseudoscalar scalar ($PS \sim$ flat) decays,
for which the shapes are obtained from our $\rho^0\pi^0$ signal MC,
and a sample of $\sigma \pi^0$ MC~\cite{ref:sigma_pi0_mc}, respectively.
We find the $PS$ level is consistent with zero, and
assign a systematic error due to the possible contribution in our signal
region of $^{+0.0}_{-5.0}\%$.
The $m_{\pi^+\pi^-}$ distribution supports the conclusion that our
signal is due to $B^0 \to \rho^0\pi^0$.

To extract the branching fraction,
we measure the reconstruction efficiency from MC and correct for 
discrepancies between data and MC in the pion identification 
and continuum suppression requirements.
The correction factor due to pion identification ($0.89$) is obtained
in bins of track momentum and polar angle from an inclusive $D^*$ control
sample ($D^{*-} \to \bar{D}^0 \pi^-$, $\bar{D}^0 \to K^+ \pi^-$). The
resulting systematic error is $\pm 3.3\%$.
For the continuum suppression requirement on $r$ and ${\mathcal R}$, 
we use a control sample $B^0 \to D^{-}\rho^+$ with
$D^{-} \to K^+\pi^-\pi^-$, $\rho^+ \to \pi^+\pi^0$,
which has the necessary feature of being a neutral $B$ decay
to ensure the $r$ behaviour is the same as that of our signal.
A correction factor of $1.15$ is obtained; the statistical error of
this control sample accounts for the largest contribution to the
systematic error, $\pm 11\%$.

We further calculate systematic errors from the following sources:
PDF shapes $^{+1.6}_{-1.5}\%$ (by varying parameters by $\pm 1 \sigma$);
$\pi^0$ reconstruction efficiency $\pm 3.5\%$ (by comparing the yields of
$\eta \to \pi^0 \pi^0 \pi^0$ and $\eta \to \gamma \gamma$ between data and MC);
track finding efficiency $\pm 2.4\%$ 
(from a study of partially reconstructed $D^*$ decays).
We use our calibration control samples to study possible effects
on the efficiency due to the $\Delta E > -0.2 \ {\rm GeV}$ requirement
and assign a $2\%$ systematic error.
The total systematic error due to possible charmless $B$ decays not 
otherwise included is $\pm 5.3\%$. 
We repeat the fit after changing the
normalisation of the fixed $B$ decay components according to the error
in their branching fractions, and obtain systematic errors from the change
in the result of $\pm 1\%$. In the case that the normalisations of 
$B$ backgrounds fixed in the fit are simultaneously increased by $1\sigma$,
 the statistical significance decreases from $3.6 \sigma$ to $3.5 \sigma$;
we interpret the latter value as the significance of our result.
Finally, we estimate the systematic uncertainty due to possible interference
with $B^0 \to \rho^\pm \pi^\mp$ by varying the $m_{\pi^\pm \pi^0}$ veto
requirement.
We find the largest change in the result (by $9.3\%$) 
when this requirement is removed, and assign this as the error.
The total systematic error is $\pm 17\%$,
and we measure the branching fraction of $B^0 \to \rho^0\pi^0$ to be
$$
  {\mathcal B}\left( B^0 \to \rho^0\pi^0 \right) =
  \left( 5.1 \pm 1.6 ({\rm stat}) \pm 0.9 ({\rm syst}) \right)
  \times 10^{-6}.
$$

In order to test the robustness of this result,
a number of cross-checks are performed. We vary the selection
on $r$ and ${\mathcal R}$. We try numerous combinations of requirements,
with efficiencies that vary between
$1.60\%$ and $2.70\%$. In all cases consistent central
values of the branching fraction are obtained.
We also repeat the analysis using a looser requirement on the lower
bound of $\Delta E$ and obtain consistent results.
Finally, we select $\rho^{\pm}\pi^{\mp}$ candidates from the 
$\pi^+ \pi^- \pi^0$ phase-space using the same continuum suppression
requirement, and measure a branching fraction for
$B^0 \to \rho^{\pm}\pi^{\mp}$ that is 
consistent with previous measurements~\cite{ref:rhopi_bfs}.

In summary, we observe the first evidence, 
with $3.5\sigma$ significance, for $B^0 \to \rho^0\pi^0$ 
with a branching fraction higher than most
predictions~\cite{ref:rho0pi0_predictions},
and a central value above the upper limit recently set 
by the BaBar collaboration~\cite{ref:rho0pi0_ul}.
Our result may indicate that some contribution 
to the amplitude is larger than expected, 
which may complicate the extraction of $\phi_2$ from the $\rho\pi$ system.

We thank the KEKB group for the excellent
operation of the accelerator, the KEK Cryogenics
group for the efficient operation of the solenoid,
and the KEK computer group and the NII for valuable computing and
Super-SINET network support.  We acknowledge support from
MEXT and JSPS (Japan); ARC and DEST (Australia); NSFC (contract
No.~10175071, China); DST (India); the BK21 program of MOEHRD and the
CHEP SRC program of KOSEF (Korea); KBN (contract No.~2P03B 01324,
Poland); MIST (Russia); MESS (Slovenia); NSC and MOE (Taiwan); and DOE
(USA).


\begin{thebibliography}{99}
\bibitem{cc}
  The inclusion of charge conjugate decays is implied 
  throughout.
\bibitem{rareb}
  Decay modes that contribute to the $\pi^+\pi^-\pi^0$ final state are
  excluded from this MC and considered separately.
\end{thebibliography}
\end{document}